\documentclass{article}

\usepackage{times}
\usepackage{graphicx} 
\usepackage{subfigure} 
\usepackage{enumitem}

\usepackage{natbib,har2nat}

\usepackage{algorithm}
\usepackage[noend]{algorithmic}

\usepackage{hyperref}

\usepackage[accepted]{icml2015}

\usepackage{url}

\usepackage{epsfig}
\usepackage{epstopdf} 

\usepackage{comment}
\usepackage{authblk}

\icmltitlerunning{Distributed DNN}

\begin{document}  

\setlength{\textfloatsep}{0.1\baselineskip plus 0.1\baselineskip minus 0.1\baselineskip}

\twocolumn[
\icmltitle{Distributed Deep Learning  
           Using Synchronous Stochastic Gradient Descent}

\icmlauthor{Dipankar Das}{dipankar.das@intel.com}
\icmladdress{Intel Parallel Computing Lab, Bangalore}
\icmlauthor{Sasikanth Avancha}{sasikanth.avancha@intel.com}
\icmladdress{Intel Parallel Computing Lab, Bangalore}
\icmlauthor{Dheevatsa Mudigere}{dheevatsa.mudigere@intel.com}
\icmladdress{Intel Parallel Computing Lab, Bangalore}
\icmlauthor{Karthikeyan Vaidynathan}{karthikeyan.vaidyanathan@intel.com}
\icmladdress{Intel Parallel Computing Lab, Bangalore}
\icmlauthor{Srinivas Sridharan}{srinivas.sridharan@intel.com}
\icmladdress{Intel Parallel Computing Lab, Bangalore}
\icmlauthor{Dhiraj Kalamkar}{dhiraj.d.kalamkar@intel.com}
\icmladdress{Intel Parallel Computing Lab, Bangalore}
\icmlauthor{Bharat Kaul}{bharat.kaul@intel.com}
\icmladdress{Intel Parallel Computing Lab, Bangalore}
\icmlauthor{Pradeep Dubey}{pradeep.dubey@intel.com}
\icmladdress{Intel Parallel Computing Lab, Santa Clara}

\icmlkeywords{Deep Learning, Distributed Machine Learning, CNN}

\vskip 0.3in
]

\begin{abstract} 
We design and implement a distributed multinode synchronous SGD algorithm, without altering hyperparameters, or compressing data, or altering algorithmic behavior.
We perform a detailed analysis of scaling, and identify optimal design points for different networks.
We demonstrate scaling of CNNs on 100s of nodes, and present what we believe to be record training throughputs.
A 512 minibatch VGG-A CNN training run is scaled 90X on 128 nodes.
Also 256 minibatch VGG-A and OverFeat-FAST networks are scaled 53X and 42X respectively on a 64 node cluster.  
We also demonstrate the generality of our approach via best-in-class 6.5X scaling for a 7-layer DNN on 16 nodes.
Thereafter we attempt to democratize deep-learning by training on an Ethernet based AWS cluster and show ~14X scaling on 16 nodes.
\end{abstract}

\section{Introduction}
\label{sec:intro}

With the efficacy of large deep neural networks well established~\cite{Schmidhuber15}, the key challenge is to train them in a reasonable amount of time, preferably hours or even minutes. The largest networks require several Exaflops of computation\footnote{VGG-A needs 33.6 GFlops per image, 43 PFlops per epoch on ImageNet-1k, and 1 ExaFlop per 25 epochs}; clearly a single node or single card implementation fails to meet this challenge. Distributed (i.e., multinode) training methods are required to address this gap.
Also, to stay relevant, deep learning must ride the scaling curve like it took advantage of increasing compute density and Moore's law in recent years.
In order to address this challenge, researchers have developed multinode/multi-card frameworks such as TensorFlow \cite{tensorflow}, FireCaffe \cite{firecaffe}, and DeepImage \cite{deepimage}.       

However, scaling synchronous Stochastic Gradient Descent (SGD) is challenging, as it is a strong-scaling problem on a small amount of work bound by compute needs for processing a minibatch of data-points (between 64-5120). Thus, many of these frameworks suffer from poor scalability, and typically do not scale beyond tens of nodes. 
Many variants of synchronous SGD such as 1-bit SGD \cite{onebitsgd}, elastic-SGD \cite{elasticsgd}, as well as asynchronous SGD \cite{adam} have been proposed for better scaling.
However, unlike these methods we do not alter hyperparameters (like minibatch or learning rate) or the algorithm, or use any compression methods, but focus on deeply understanding the vanilla SGD algorithm and scaling it.
 
Our approach is to systematically develop detailed system balance equations, and solve them to obtain limits for performance; we also find optimal design points for Xeon-based multinode training frameworks. We examine data-parallelism and model-parallelism, and propose a new algorithm for hybrid data- and model-parallelism. Based on this analysis we identify which strategy is best suited for different layers of a neural network. 
   
However, before we build a distributed deep-learning training framework, we must first optimize a single node implementation to the highest possible efficiency. Again, we take an analytic approach to study the balance between computation and memory bandwidth in order to obtain optimal cache blocking strategies. 
Specific to x86-based architectures, we present details of threading, register blocking, as well as instruction sequence of innermost loops, and analyze how these techniques enable a deep learning framework to achieve high single-node efficiency.

We note that the analysis in this work is generic and applicable, to the best of our understanding, to other non-x86-based deep learning systems as well (including those using GPUs and accelerators). 

Combining a highly efficient single node implementation and an optimized multi-node scaling strategy, we present the best time-to-train for several CNNs used for image recognition, and break all published training throughput records using Xeon based systems. We achieve efficiency of \~90\% for several convolutional layer operations and \~70\% for fully-connected layers on Intel Xeon E5-269Xv3 machines for a wide variety of neural networks. Our multinode algorithms and tuned implementations help us significantly surpass best published scaling efficiency for a wide gamut of neural networks.   

The rest of the paper is organized as follows: in section~\ref{sec:compute} we discuss balance equations for the convolutional and fully-connected {\it compute} layers and in section~\ref{sec:comms} for the {\it communication patterns}. Section~\ref{sec:software} describes the components of our software framework and their design. We present detailed single-node and multi-node performance results for two well-known deep learning topologies -- OverFeat~\cite{overfeat} and VGG-A~\cite{vgg} -- on the Cori system~\footnote{http://www.nersc.gov/users/computational-systems/cori/} and AWS~\footnote{https://aws.amazon.com/}.

\section{Optimizing Computation in Neural Network Training}
\label{sec:compute}

We can view the computation of neural network training as a task graph where each node represents a block of computation (typically one layer), and edges represent data-flow (usually a multi-dimensional tensor), which define data-dependencies between nodes.
It is critical to understand the compute and memory-bandwidth needs of nodes, and identify optimal threading, cache-blocking, vectorization, and register-blocking strategies.

\subsection{Compute Patterns}
 The compute heavy convolution and fully-connected layers take two $k+2$-dimensional tensors as input and produce a $k+2$-dimensional output tensor.
Here $k$ is the number of dimensions of a feature map, or kernel.
The additional two dimensions depend on the type of data: for tensors containing inputs and outputs or gradients of inputs and outputs, they represent the {\it minibatch} and {\it feature} identifier; for tensors containing weights and gradients of weights, they represent a pair of {\it input-output features}.
 
The compute operations for forward propagation, backpropagation and determining weight gradient are identical $2k+3$-dimensional loops.
As an example, consider a 2-D convolution forward propagation operation, where the input is a 4-D tensor over minibatch, input feature map (ifm) identifier, output feature map height ($out_h$) and width ($out_w$) respectively.
The weight tensor is another 4-D tensor over input feature map, output feature map (ofm), kernel width ($k_w$) and kernel height ($k_h$). In the rest of the paper, we refer to $kernel$ and $weights$ interchangeably. Algorithm~\ref{alg:fp} describes the forward propagation operation. The variable $s$ represents the stride for the convolution.

\begin{algorithm}[t]
\caption{Forward Propagation}
\label{alg:fp}
\begin{algorithmic}[1]
\small
\FOR{$i_0 \in 0,\dots,minibatch$}
	\FOR{$i_1 \in 0,\dots,ifm$}
    	\FOR{$i_2 \in 0,\dots,ofm$}
        	\FOR{$i_3 \in 0,\dots,out_h$}
            	\FOR{$i_4 \in 0,\dots,out_w$}
                	\FOR{$i_5 \in 0,\dots,k_h$}
                    	\FOR{$i_6 \in 0,\dots,k_w$}
                        	\STATE{$output[i_0, i_1, i_3, i_4] +=$}
                            \STATE{$input[i_0, i_1, i_3 * s + i_5 - 1, i_4 * s +i_6 - 1] * wts[i_1, i_2, i_5, i_6]$}
                        \ENDFOR
                     \ENDFOR
                  \ENDFOR
              \ENDFOR
          \ENDFOR
      \ENDFOR
\ENDFOR
\end{algorithmic}                            
\end{algorithm}

A fully-connected layer can similarly be written as a special case of this 7-nested loop when $k_h$, $k_w$, $out_h$, $out_w$ are all 1.
Moreover other operations such as backpropagation and weight gradient computation have identical loops with different multiply and accumulate operations.

Backpropagation:\\
$grad\_input[i_0, i_1, i_3 * s  + i_5 -1, i_4 * s + i_6 -1] +=$\\
$grad\_output[i_0, i_2, i_3, i_4] * wts[i_1, i_2, i_5, i_6]$

Weight-gradient Update:\\
$grad_wts[i_1, i_2, i_5, i_6] += $\\
$input[i_0, i_1, i_3 * s  + i_5 -1, i_4 * s + i_6 -1] *$\\
$grad\_output[i_0, i_2, i_3, i_4]$

This similarity between all the operations implies that the memory access pattern for all three operations is identical, and therefore, a cache blocking strategy for one of them should apply to others as well.

\subsection{Cache Blocking}
Unless the activations and weights completely fit within the CPU cache hierarchy (which is often not the case), the loop over $i_3$ reads $out_h * out_w$ output activations, $(out_h * s + k_h - 1)*(out_w * s + k_w - 1)$ input activations (which we denote $in_h * in_w$), and $k_h * k_w$ weights from external memory (i.e., DRAM). The loop performs $k_h * k_w * out_w * out_h$ multiply-and-accumulate operations. 
Therefore, the Bytes to FLOPs (floating-point operations) ratio is:\\ 
$B/F = size_{data} * (out_w * out_h + in_w * in_h + k_w * k_h)$
$/(2 * k_w * k_h * out_w * out_h)$.  

For a convolutional layer with $12*12$ output, $3*3$ kernel, 512 input feature maps and 1024 output feature maps (such as C5 in OverFeat-FAST), the B/F ratio is 0.54; typically the system B/F ratio is less than 0.08.
On the other hand, if all data (inputs, outputs, weights) fit into the cache hierarchy, the B/F ratio reduces because with a one-time read from DRAM, all 7 loops can be computed at one go:\\
$B/F = size_{data} * (minibatch * ofm * out_w * out_h + minibatch * ifm * in_w * in_h + ifm * ofm * k_w * k_h)$
$/(2 * minibatch * ofm * ifm * k_w * k_h * out_w * out_h)$.  

Now, the best achievable B/F ratio for C5 in OverFeat-FAST is 0.003. Clearly, the capacity (i.e., size in bytes) of the cache hierarchy determines the range of B/F ratios. Given the capacity, we formulate the cache blocking problem as a constrained minimization problem ($s_1, s_2, \dots s_{k+2}$ are strides):

$BS = size_{data} * (b1_0 * \ldots b1_{k+2} + b2_0 * \ldots b2_{k+2} + b1_0 * b2_0 * (b1_2*s_1+b2_2-1)\ldots (b1_{k+2}*s_{k+2} + b2_{k+2}-1))$\\
$CPB = 2 * b1_0 * b1_2 * \ldots b1_{k+2} * b2_1 * b2_2 * \ldots b2_{k+2}$\\
$B/F = BS/CPB$\\
$\forall_i~find~b1_i,~b2_i$ that minimize $B/F$, s.t. $BS < Size_{cache}$

Here, $BS$ is the size of the block residing in on-chip memory, $CPB$ is the amount of compute to perform on the block, and $Size_{cache}$ is the size of the on-chip memory/cache (with due consideration for double buffering).

We observe that traversing along consecutive blocks in any dimension, results in memory reuse and therefore, better B/F ratios.
For example, consider the scenario where block size along the loop $i_3$ is 1, and we traverse along the {\it height}-dimension in a 2-D convolution with stride=1.
Here we need to read in only one row of $ifm$ for each row of $ofm$, instead of $k_h$ rows of $ifm$.
Similarly, traversing along the $ifm$ dimension precludes reading the output-block.

We write a multithreaded program to perform a brute-force state space search over all values of loop iterators in order to find the minimum B/F ratio for different 2-D convolutional layers, given a limit on the cache size. 
Additionally, to compute $BS$, we constrain one of the dimensions to be a multiple of SIMD-width or warp-size since all modern high-performance processors use some form of SIMD operation.
Hence the sub-problem to be solved must have one dimension (preferably the output dimension) set to a multiple of SIMD-width.

We find that with 128 KB of cache per thread in modern Xeon CPUs, a B/F ratio of $\le$ 0.04 can be maintained for most convolutional layers even for a minibatch size of 1.

\begin{algorithm}[t]
\caption{Generic Optimized Forward Propagation}
\label{alg:ofp}
\begin{algorithmic}[1]
\small
\FOR{$i_0 \in 0,\dots,minibatch$}
	\FOR{$i_1 \in 0,\dots,ifm/SW$}
    	\FOR{$i_2 \in 0,\dots,ofm/SW$}
        	\FOR{$i_3 \in 0,\dots,out_h/RB_h$}
            	\FOR{$i_4 \in 0,\dots,out_w/RB_w$}
                	\FOR{$rb_h \in 0, \dots, RB_h$}
                    	\FOR{$rb_w \in 0,\dots, RB_w$}
                        	\STATE{$reg = rb_h * RB_w + rb_w$}
                        	\STATE{$out_y = i_3 * RB_h + rb_h$}
                            \STATE{$out_x = i_4 * RB_w + rb_w$}
                        	\STATE{$vout[reg]$ $=$ $LOAD(output[i_0][i_2][out_y][out_x])$}
                         \ENDFOR
                    \ENDFOR
                	\FOR{$i_5 \in 0,\dots,SW$}
                		\FOR{$i_6 \in kh_{start},\dots,kh_{end}$}
                    		\FOR{$i_7 \in kw_{start},\dots,kw_{end}$}
                            	\STATE{$vwt = LOAD(wts[i_1 * SW + i_5][i_2][i_6][i_7][0])$}
                            	\FOR{$i_8 \in 0,\dots,RB_h$}
                                	\FOR{$i_9 \in 0,\dots,RB_w$}
                                    	\STATE{$reg = i_8 * RB_w + i_9$}
                                    	\STATE{$out_y = i_3 * RB_h + i_8$}
                                        \STATE{$out_x = i_4 * RB_w + i_9$}
                                        \STATE{$inp_y = out_y* stride + i_6 -1$}
                                        \STATE{$inp_x = out_x * stride + i_7 -1$}
                                    	\STATE{$vout[reg] = VFMA(vout[reg],$ $bcast (input[i_0][i_1][out_y][out_x][0]))$ $, vwt)$}
                                	\ENDFOR
                                \ENDFOR
                        	\ENDFOR
                     	\ENDFOR
                     \ENDFOR
                     \FOR{$rb_h \in 0, \dots, RB_h$}
                    	\FOR{$rb_w \in 0,\dots, RB_w$}
                        	\STATE{$reg = rb_h * RB_w + rb_w$}
                        	\STATE{$out_y = i_3 * RB_h + rb_h$}
                            \STATE{$out_x = i_4 * RB_w + rb_w$}
                        	\STATE{$STORE(vout[reg], output[i_0][i_2][out_y][out_x])$}
                         \ENDFOR
                    \ENDFOR
                  \ENDFOR
              \ENDFOR
          \ENDFOR
      \ENDFOR
\ENDFOR
\end{algorithmic}                            
\end{algorithm}

\subsection{Data Layout and Vectorization}
In addition to cache blocking, we need to vectorize the operations and perform register blocking as well. 
A fully optimized vectorized forward propagation operation is presented in Algorithm \ref{alg:ofp}.
It contains a 10-nested loop with cache blocking blocking along $ifm$, and $ofm$, and register blocking along $output_h$, and $output_w$ dimensions.
For this blocked loop structure we lay out data so that access in the innermost loops is as contiguous as possible. 
This results in better utilization of cache lines (and hence bandwidth) and improves prefetcher performance.

In this work we lay out all data, including activations and weights with the innermost dimension over groups of SIMD-width ($SW$) output feature maps. That is we lay out the different data structures as:\\ 
 and gradient of activations: $N \times C \times H \times W \rightarrow N \times (C/SW) \times H \times W \times SW$\\
Weights and gradients of weights: $IFM \times OFM \times KH \times KW \rightarrow IFM \times (OFM/SW) \times KH \times KW \times SW$\\
Transpose-weights: $IFM \times OFM \times KH \times KW \rightarrow OFM \times (IFM/SW) \times KH \times KW \times SW $

Here, N stands for minibatch, C stands for feature-maps, H for feature map height, W for feature map width, IFM for input feature maps, OFM for output feature maps, KH for kernel-height and KW for kernel-width.
This layout also enables vectorization of operations, such that the multiply-and-accumulate in Algorithm \ref{alg:fp} can now be replaced by a broadcast and vector fused-multiply and add (Algorithm \ref{alg:ofp}).
\vspace{-4mm}
\subsection{Register Blocking}

The aim of register blocking is two-fold: firstly it improves the ratio of vector fused multiply and add (VFMA) operations to that of load/store operations.
Secondly, a sequence of $reg$ consecutive VFMA instructions is needed to hide the latency of these instructions.
The latency for a VFMA operation on the Xeon CPU core is 5 cycles, and a Xeon CPU core can execute 2 VFMA instructions per cycle.
Hence in order to completely hide the latency of VFMA instructions we should have a register block size of at least 10. 
Hence $15 \ge RB_h * RB_w \ge 10$, as we need one register to store the weights.

In Algorithm \ref{alg:ofp} we illustrate a 2-D register block for forward propagation.
Since a Xeon core can perform 2 Loads per cycle, 2 VFMAs per cycle, and 1 store per cycle, the cycles spent on load/store instructions ($LS$) and VFMA instructions ($FMA$) for the inner loop (line 5-29 in Algorithm \ref{alg:ofp}) can be computed as:

$LS = (RB_h * RB_w + SW * (kh_{end} - kh_{start}) * (kw_{end} - kw_{start}))/2 + RB_h * RB_w$

$FMA = (SW * (kh_{end} - kh_{start}) * (kw_{end} - kw_{start}) * RB_h * RB_w)/2$

In practice $RB_h$ is often 1 for forward propagation, since most feature map width are $\ge 12$ for CNNs.
For $RB_w = 12$ and $kh_{start} = kh_{end} = 0$ and $kw_{start} = kw_{end} = 3$ (as in case of OverFeat-FAST C5 layer), we compute the efficiency to be: 88\%.
The loop for backpropagation is similar with blocking along $inp_w$, $inp_h$ and $ifm$, instead of $out_w$, $out_h$, and $ofm$.

Unlike forward and backpropagation, weight gradient computation has the weight (-gradient) kernel as the output.
These kernels are often small (3x3, 5x5, 7x7, 11x11), and even two dimensional blocking will only yield a theoretical peak efficiency of 75\% for a 3x3 kernel.
Hence there is often a need to perform blocking along the $ifm$ dimension as well.
Indeed we propose using specific tailored strategies for different kernel sizes.
\vspace{-4mm}
\begin{itemize}[noitemsep,nolistsep]
\item 3x3 kernel: register block with one row (3 SIMD-elements) of 4 consecutive kernels along the input-feature-map dimension.
\item 5x5 and 7x7: register block with one row of 2 consecutive kernels along the input-feature map dimension.
\item 11x11: one dimensional register block along the kernel-width dimension.
\end{itemize}

\subsection{Threading and Work Partitioning}
We perform fine grained partitioning of work across threads. For the forward and backpropagate operations, we partition the work across multiple minibatches into jobs, each for one row of the output/input across $SW$ output/input features. 
These jobs are then equally distributed across the different threads (iterations in lines 1-4). 
For weight update, we treat weight kernels for $SW$ input- and output-feature pairs to be the basic unit of work, and these jobs are subsequently distributed across multiple threads.
In case the number of jobs created in this way is low (like for C1 layers), we additionally partition the problem along the minibatch dimension, and then privtize and reduce weight gradient computation.

\section{Optimizing Communication}
\label{sec:comms}
In this work we perform strong scaling of the synchronous minibatch stochastic gradient descent algorithm. 
We scale the computation for one iteration across multiple nodes, such that the multi-threaded, and multi-node parallel implementation is equivalent to a single-node single-threaded serial implementation. 
We present a detailed theoretical analysis of computation and communication balance equations, and determine strategies for work partitioning between nodes.   

\subsection{Data Parallelism}
We analyze the algorithmic computation-to-communication balance of data parallelism, wherein the work in an iteration is partitioned across the minibatch. 
For our analysis we consider a butterfly-reduce operation.

Consider a convolutional layer with $ofm$ output feature maps each of size: $out_w * out_h$ (width and height), $ifm$ input feature maps, $s$ stride, and kernel of size $k_w * k_h$. 
The amount of computation (for $MB_{node}$ data points assigned to a node) in the number of FLOPS in this layer for forward, backward, and weight gradient computation steps is:\\ 
       $Comp = 3 * 2 * MB_{node} * ifm * ofm * k_w * k_h * out_w * out_h$

Similarly, we estimate the total communication per iteration for a data-parallel approach. 
In each iteration, a node sends partial weight gradients to other nodes, and receives updated weights from them. Therefore, the total communication volume is:\\
	$Comm = size_{data} * ifm * ofm * k_w * k_h * (2 - overlap))$

$overlap$ is the amount of overlap that the software achieves between send and receive operations. 
Assuming floating point data representation, and $overlap = 1$ of sends and receives, the algorithmic communication-to-computation ratio for data parallel implementation of a single layer is:\\
       $comp\_comm = 1.5 * out_w * out_h * MB_{node}$

We observe that the algorithmic computation-to-communication ratio does not depend either on the kernel size or number of input and output feature maps or stride, but solely on the size of the output feature-map and the number of data-points assigned per node.

An advantage of data parallelism is that it can be overlapped with computation of current and previous layers.
Moreover, the weight update function needs weight gradient values which are available immediately after the backpropagation of a given layer {$k$} and the updated weights are not needed until before the forward propagation step for layer {$k$} in the next iteration. 
Therefore, we estimate the scalability of layers $L_0, L_1, L_2, ..., L_{k-1}$, as follows:

$ocomp_i = \sum_{j < i} comp_j + comp_i/3$\\
$ocomms_i = \sum_{j <= i} comms_j$\\
$bubble_i = ocomms_i/comm_{sys} - ocomps_i/comps_{sys}$

Scaling efficiency is then estimated as the ratio of two terms:
$(\sum_{i<k} bubble_i + (\sum_{i<k} comp_i)$ and $(comp_{sys})/((\sum_{i<k} comp_i)/comp_{sys})$\\

The best compute-communication overlap implies that $\forall_{i>0} bubble_i = 0$, as we cannot avoid the communication bubble between the weight-gradient update and forward propagation steps of the first layer $L_0$. 
It is notable that the first layer need not perform backpropagation and can only perform weight gradient computation.
Moreover, we perform weight gradient computation before backpropagation in order to allow for some more computation to overlap communication (which explains the $comp_i/3$ term).

We note that in typical convolutional networks (and otherwise), the size of feature maps keeps on reducing monotonically with increase in layer-id. 
Also, we noted earlier that the algorithmic compute-to-communication ratio depends on only the feature map size; thus, if the communication of layer $l$ cannot be completely overlapped with compute, then the communication of layer $l+1$ cannot be overlapped as well.
Therefore, to estimate the best possible compute-communication overlap we check if $bubble_{k} < 0$, where layer $L_k$ is the last layer in the data-parallel regime.
For convolutional neural networks, this is usually the last convolutional layer. The number of nodes (N) to which the algorithm scales is: \\ 
$N \le minibatch * (comms_{sys}/comp_{sys}) * (ocomp_k/ocomms_k)$ 
The algorithmic computation-to-communication ratio convolutional layers of OverFeat-FAST and VGG-A are 208, and 1456 respectively.

We examine the smallest number of data-points per node for a training run of OverFeat-FAST and VGG-A on two platforms: dual-socket 16-core Xeon E5-2698v3 + FDR Infiniband, and 2-socket 9-core 2.9 GHz Xeon E5-2666 v3 + 10GigE Ethernet (Table \ref{dataparallelism}). 
Based on this, we estimate the scaling for data parallel parts of a 256 minibatch training run.
We estimate that the convolutional layers can be scaled to 128 nodes for OverFeat-FAST and 256 nodes for VGG-A.
Note that in a CNN there are several fully connected layers which do not scale much in practice as compared to convolutional layers.

\begin{table}
\centering
\caption{Theoretical Scaling of Data Parallelism. Minimum number of data points per node as well scaling of a 256 minibatch problem for convolutional laters.}
\label{dataparallelism}
\begin{tabular}{|c|c|c|}
\hline
 & 2s9c E5-2666v3 & 2s16c E5-2698v3 \\
 & + 10Gbps Ethernet &  + 56Gbps FDR \\
\hline
Comp-to-comms & 1336 & 336\\
\hline
OverFeat-FAST & 3 (86) & 2 (128)\\
\hline
VGG-A & 1 (256) & 1 (256)\\
\hline
\end{tabular}
\end{table}

\subsection{Analyzing Model Parallelism}

We first consider a simple model parallel approach where each node operates on a part of the model of size: $ifm_b * ofm_b$ input- and output-feature maps. 
In this case, the computation for the forward pass, backward pass, or weight-gradient update is given as:

$Computation = 2 * ifm_b * ofm_b * kernel_w * kernel_h * output_w * output_h * minibatch$

For the forward pass the amount of data received is:

$comms_{recv} = size_{data} * ifm_b * input_w * input_h * minibatch * (ifm/ifm_b - 1)$

The amount of data sent out by the previous layer is:

$comms_{send} = size_{data} * ifm_b * input_w * input_h * minibatch$

Hence the total volume of communication data is: $size_{data} * ifm * input_w * input_h * minibatch$

Hence the time taken for a forward pass with no compute and communication overlap for a given layer is:

$Computation/SysFlops + (comms_{recv} + comms_{sent})/CommBW + SWlat$

It is notable that the communication-bandwidth is dependent on the message size, and along with the impact of software overheads, the performance of model parallel communication pattern falls sharply with decrease in size of the feature map and the number of features.

A question of interest is to determine when is model parallelism preferred to data parallelism.
We do a simplified analysis to compare the amount of communicated data in each method. Model parallelism is better if:

 $size_{data} * ifm * ofm * kernel_w * kernel_h * (1 + (1 - overlap)) > size_{data} * ifm * input_w * input_h * minibatch$

We can simplify this further to:
 $ ofm * kernel_w * kernel_h * (2-overlap) > input_w * input_h * minibatch$

In convolutional layers $ofm$ is typically less than 1024 and $kernel_h$ and $kernel_w$ are 3, or 5, while $input_h$ and $input_w$ are greater than 14, and minibatch size $>$ 64.
For this case, only for a large kernel size and small minibatch does model parallelism become better.
For fully connected layers, where $kernel_w, kernel_h, input_w and input_h$ are 1 whenever $ofm > minibatch$ model parallelism is better than data parallelism. 
This is typically the case for most fully connected layers, unless we have large minibatches ($> 5000$) as in case of ASR networks.

\subsection{Analyzing Hybrid Parallelism}

Beyond vanilla data- and model-parallelism we explore a hybrid scheme.
One may view data-parallelism as partitioning work along the "minibatch" dimension and model parallelism as partitioning along the "feature map" dimension. 
Clearly we can partition work along both "minibatch" as well as "feature map" dimensions.

For this we partition nodes into node groups, such that nodes within a group follow a model-parallelism regime while corresponding nodes across node groups follow a data-parallelism regime.

In this scheme the minibatch is partitioned into $G$ groups each containing $N/G$ nodes and responsible for $mb_{group} = minibatch/G$ data-points. 
Model parallelism on this subgroup, for forward as well as back-propagation, leads to an exchange of $comms_model$ amount of data:
$comms_{model} =  2 * size_{data} * ifm * input_w * input_h * mb_{group}$

Communication due to data-parallelism (send and receive weights for a $1/G^{th}$ fraction of the weights): 
$comms_{data} = size_{data} * ofm * ifm * kernel_w * kernel_h * (2-overlap) / G$

By changing $G$ and hence $mb_{group}$ we can balance the volume of communication for data and model parallelism. 
Recall that for data parallelism we have the option to overlap the same across all previous layers, thereby hiding time taken for communication. 
Trading off some model-parallelism for data-parallelism also helps in increasing message sizes, thereby improving network performance. 
None the less it is of interest to ask if hybrid data- and model-parallelism can yield any benefits in terms of overall communication volume:

$comms_{hybrid} = 2 * size_{data} * ifm * input_w * input_h * mb_{group} + size_{data} * ofm * ifm * kernel_w * kernel_h * (2-overlap) * (G/N)$    {\bf for $G>1$}\\
$2 * size_{data} * ifm * input_w * input_h * minibatch$ {\bf for $G=1$}

We can find the minimum value for the total communication volume by differentiating the expression for overall communication volume over G, and then solving for $d(comms_{hybrid})/d(G) = 0$.
For a fully connected layer, and assuming FP32 data-type (and no overlap), we can find the optimal point by solving: 

$d(8 * ifm * (minibatch/G + ofm * G/N))/d(G) = 0$\
Or, $-minibatch/G^{2} + ofm/N = 0$, hence $G = \sqrt(N * minibatch/ofm)$ or $G=1$.
For a layer with ofm=4096, minibatch=256, N=64, we have G = 3, and the communication volume is: $8 * ifm * 213$. For the pure model parallelism case of G=1, the communication volume is $8 * ifm * 256$. 
Clearly hybrid parallelism offers better overall communication volume than data- or model-parallelism.

As far as parallelism is concerned we believe that hybrid parallelism (with data- and model-parallelism as special cases) is often sufficient approach to parallelizing neural network computation.
Indeed one can argue that while hybrid parallelism partitions work along both the minibatch and features (possibly 2) dimensions, partitioning work across other tensor dimensions will always be sub-optimal and should be used only if enough parallelism is not afforded by partitioning across these dimensions.

\subsection{Deep Learning Communication Primitives}
\begin{figure}[ht]
\label{fig:comm_prim1}
     \centering
			\includegraphics[width=.5\columnwidth]{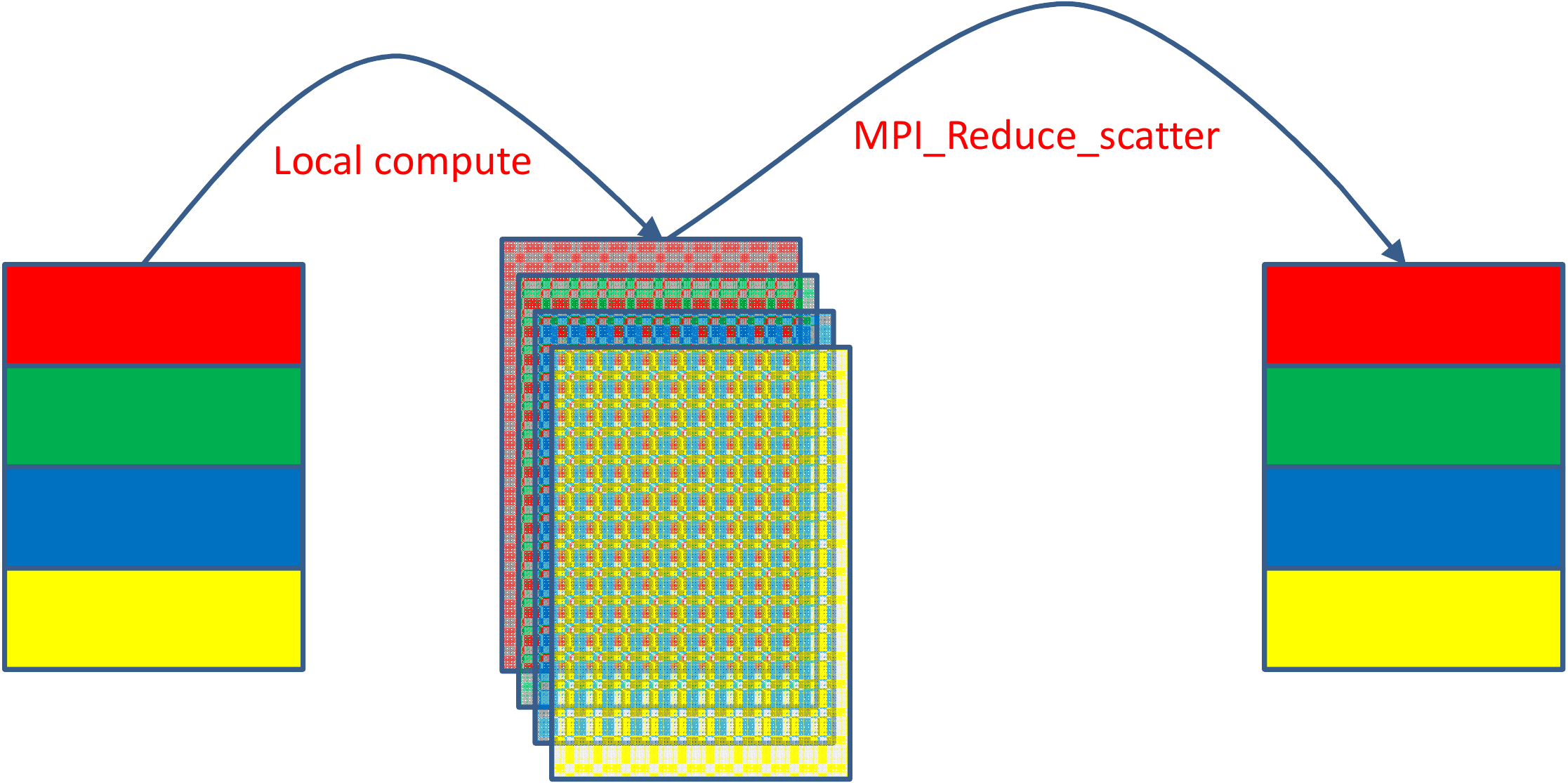}
			\caption{Communication Primitive: Part-Reduce}
\end{figure}

\begin{figure}[ht]
\label{fig:comm_prim2}
     \centering
			\includegraphics[width=.5\columnwidth]{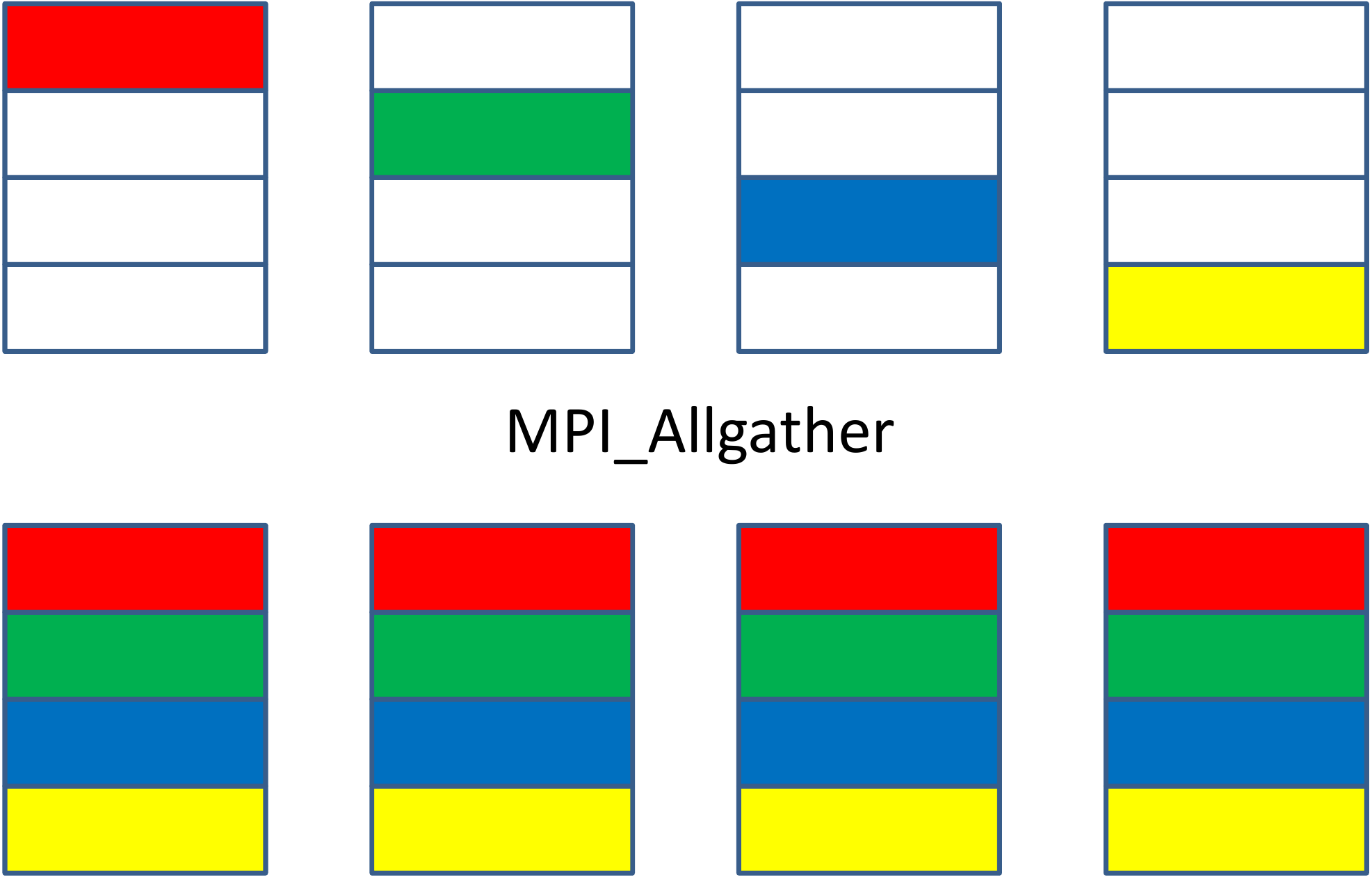}
			\caption{Communication Primitive: Part-Broadcast}
\end{figure}

The hybrid parallel approach can be implemented using two simple multi-node data transfer operations which we call: {\it part-reduce} and {\it part-broadcast}. 
Given a group of nodes $N_g$, and a tensor $\tau$, the part-reduce operation performs reduction over partial $\tau$ computed locally on each node of $N_g$ and then scatters the reduced $\tau$ to all the nodes in $N_g$. The two steps of reduction and scatter can be fused and best represented by {\it MPI\_Reduce\_scatter()} as shown in Figure~\ref{fig:comm_prim1}. In data parallelism, this primitive is required in between local weight gradient computation and performing SGD to calculate updated weights and in model parallelism, this primitive is required in forward propagation before partial activation of one layer is used as input for next layer. Similarly, in part-broadcast operation each node of $N_g$ broadcasts its locally owned strip of $\tau$ to all other nodes of $N_g$ and can be best represented by {\it MPI\_Allgather()} operation as shown in Figure~\ref{fig:comm_prim2}. This primitive is used to populate updated weights to all the nodes of $N_g$ after performing distributed SGD in the back propagation phase. It is also used to construct full gradient with respect to inputs in the back pass of model parallelism.

\section{PCL-DNN Software Framework}
\label{sec:software}

The PCL-DNN software framework consists of three primary modules: data handling, optimized compute library of core CNN/DNN functions optimized for x86 and an optimized MPI-based communications library to enable PCL-DNN execute on a large-scale distributed system. The data handling module functions as the data layer in our framework, ensuring a continuous stream of input data (e.g., images, speech) to the optimized compute library. This library drives the training or scoring process for a given application, as required. It executes the various computations -- forward propagation, backpropagation and weight updates -- on the underlying hardware.

The main role of the data handling module is to pre-process input data that the compute library uses for its functions. An important requirement we place on this module is that it must not become the bottleneck to the overall throughput of the framework, either for training or classification. That is, it must ensure continuous availability of pre-processed data to the compute library. Further, it must not compete with the latter for hardware resources (i.e., threads and cores) to perform its job. To meet these requirements, we make two design choices: one, the data handling module executes on a dedicated hardware thread and two, we access the disk via the Linux File I/O interfaces, relying on Lustre File System (LFS) to provide high-performance disk access. 

The compute library processes the input data in a layer-by-layer manner according to the specified network topology. It consists of and executes primitives for convolutional and fully-connected operations with high-efficiency on the x86 architecture using AVX2 vector instructions. The library supports convolutions with filters of various sizes (e.g., 3x3, 5x5, 11x11) with different strides. To perform the compute associated with fully-connected layers, the optimized library implements highly efficient block-SGEMM functions as well as data layout transformations.

The optimized communications library, similar to the data handling module, executes on a dedicated thread, thereby ensuring clear separation between compute and communication resources on the underlying hardware. Similar to the data handling module, it is critical that the communications library not be a bottleneck to the compute library. To this end, it performs two critical tasks: one, overlap communications during backpropagation due to hybrid parallelism with compute in the forward propagation, including message reordering and ensuring contiguous access to message buffers; two, provide a lock-free command queue~\cite{swoffloadsc2015} that enables the compute library to submit communication commands in a non-blocking manner (i.e., submit-and-forget).

\section{Experimental Results}
In this section we present performance results. We use the Cori phase I system for the for both single and multi-node experiments. This is a Cray XC machine with 1630 Intel Xeon E5-2698v3 HSW dual socket CPUs with 16 cores (supporting upto 32 threads) per socket and 128 GB of memory per node. This has Cray Aries high speed "dragonfly" topology interconnect.

\subsection{Single-node Performance}
\begin{figure}[ht]
     \centering
			\includegraphics[width=.8\columnwidth]{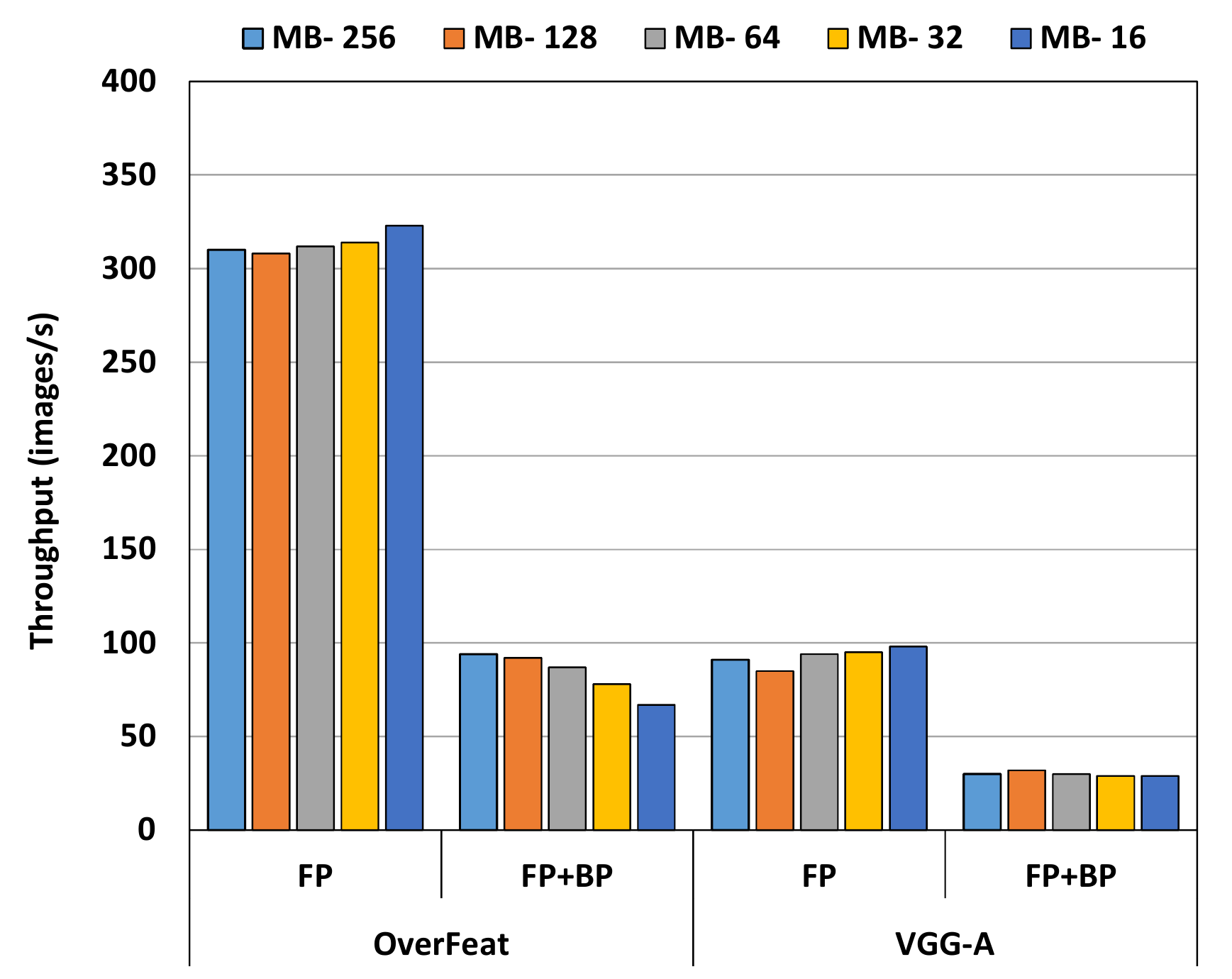}
			\caption{Single node performance and mini-batch scaling for OverFeat and VGG-A}
			\label{fig:single}
\end{figure}

We report the single-node performance of the PCL-DNN framework for the three topologies listed above. We show results for both scoring (labeled FP in the figure) and training (FP+BP in the figure) across five minibatch sizes, 16, 32, 64, 128 and 256. From Figure~\ref{fig:single}, we observe that our framework delivers approximately 315 and 95 images/s for Overfeat-FAST and VGG-A, respectively, for scoring; for training, the throughputs are approximately 90 and 30, respectively. We also observe that the throughput PCL-DNN delivers across minibatch sizes remains nearly the same for the largest topology VGG-A for both scoring and training. For OverFeat, which is approximately 3x smaller than VGG-A, we observe that the throughput for the smaller minibatch sizes (i.e., 16 and 32) is lower, than that of the largest minibatch (256) for training, whereas there is no significant variation in the throughput for scoring. We attribute lower training throughput for smaller minibatch sizes to load imbalance on the system.  These results clearly indicate that PCL-DNN executes with high-efficiency on the x86 architecture even for small minibatch sizes -- a fact that is critical to achieve high scaling efficiency across a large cluster.

\subsection{Scaling Results}
\begin{figure}[ht]
     \centering
			\includegraphics[width=.8\columnwidth]{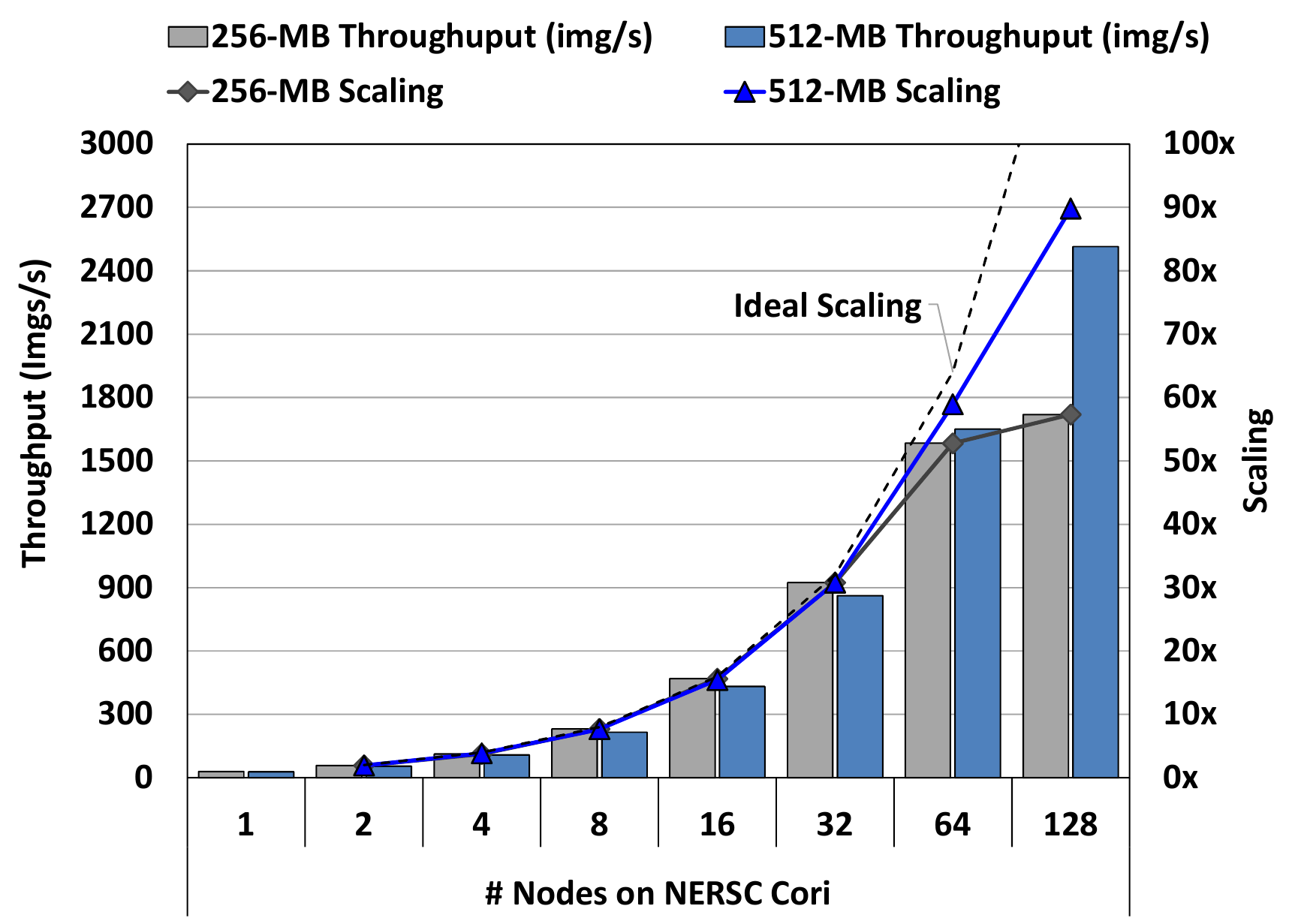}
			\caption{VGG-A performance scaling on the NERSC Cori machine up to 128 nodes (each node- Xeon E5-2698v3 HSW CPU with 16x2 cores)}
			\label{fig:vgga_scaling}
\end{figure}

\begin{figure}[ht]
     \centering
			\includegraphics[width=.8\columnwidth]{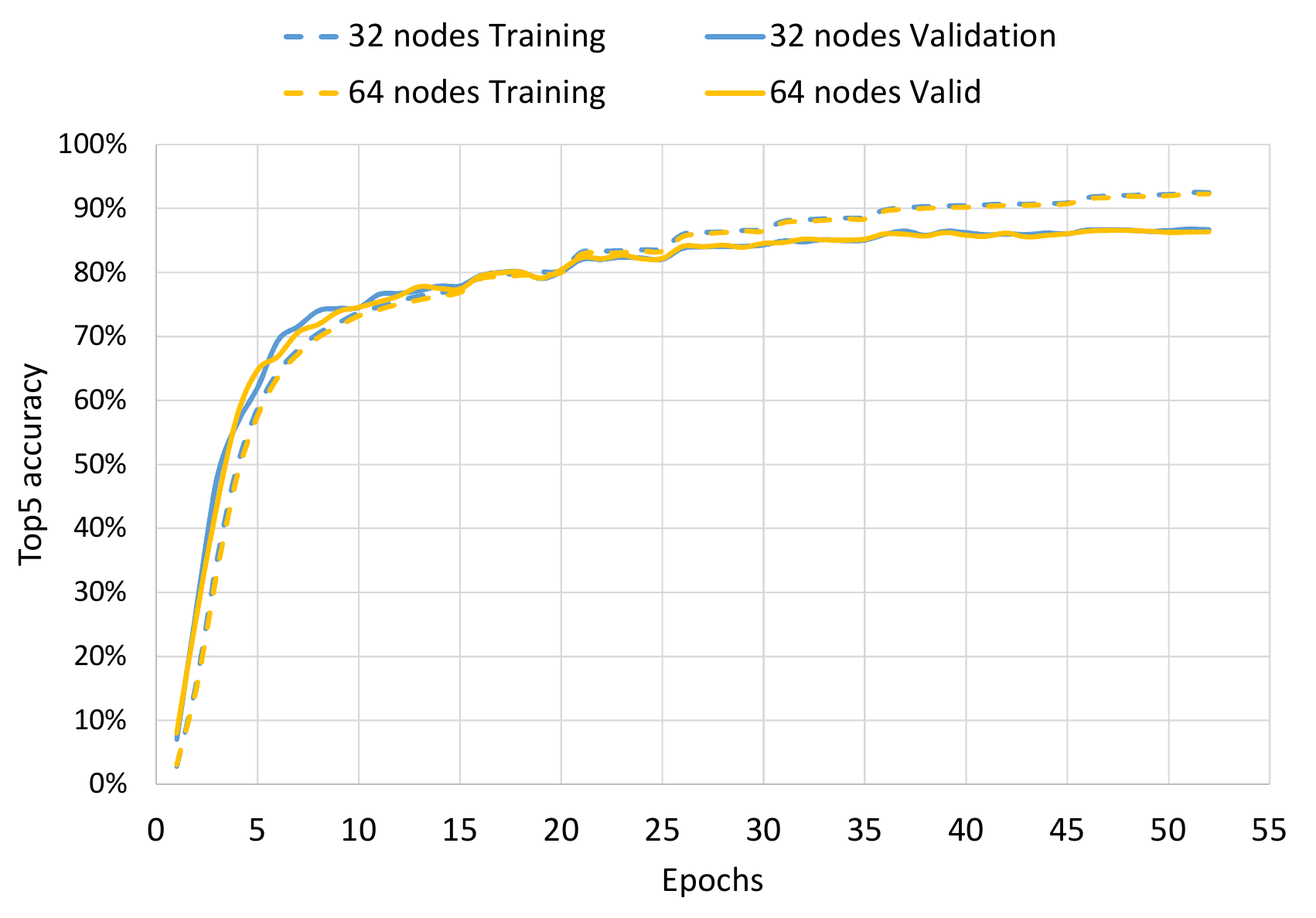}
			\caption{Top5 accuracy for VGG-A training on 32 and 64 nodes}
			\label{fig:vgga_acc}
\end{figure}

PCL-DNN delivers best till date scaling performance for deep learning applications, without using specialized HW (accelerators, NW elements, fabrics \dots). We present scaling performance for the VGG-A on the NERSC Cori phase I cluster running upto 128 nodes.
\\
Fig~\ref{fig:vgga_scaling} show VGG-A performance as we scale from 1 to 128 CPU nodes on the NERSC Cori machine. For minibatch sizes of 256 and 512, PCL-DNN scales almost linearly with the number of node. For 512 minibatch, on 128 nodes PCL-DNN scales by $90X$, with a throughput of 2510 images per second this corresponds to scaling efficiency of $70\%$ at 128 nodes. This significantly brings down the training time,\emph{ to under 10 minutes per epoch for the Imagenet-1K dataset\cite{imagenet}}. Further we show that even for a smaller minibatch of 256, PCL-DNN scaled well upto 64 nodes with a efficiency of $82\%$.
Fig~\ref{fig:vgga_acc} shows the training of the VGG-A network on 32 and 64 nodes. Since we parallelize SGD retaining its synchronous nature, and there are no hyperparameter changes, the convergence of the distributed algorithm is identical to the single node version, and the behavior for different distributed versions is identical. We see this in Fig~\ref{fig:vgga_acc}, where the Top5 validation and training accuracy overlap for both 32 node and 64 node runs.


\subsection{Scaling on Cloud}

This section presents the performance and scaling of PCL-DNN on AWS EC2. The purpose of these experiments is to demonstrate the applicability of our performance optimizations in multi-tenant cloud environments that are not as high performing as dedicated HPC clusters. Our experiments were performed on a cluster of 16-node c4.x8large RHEL 7.1 instances. Each instance consists of 2-socket 9-core 2.9 GHz Xeon E5-2666 v3 with 60 GB memory. Instances are connected via 10 Gigabit Ethernet. Unlike Cori, a dedicated bare-metal HPC cluster, CPU and network resources in EC2 are virtualized incurring higher overheads. 


We enabled SR-IOV network virtualization support (aka 'enhanced networking' by AWS). Further, we dedicated a core for handling interrupt requests for the network transmit and receive queues, resulting in 30\%-40\% better network performance. 

\begin{figure}[ht]
     \centering
			\includegraphics[width=\columnwidth]{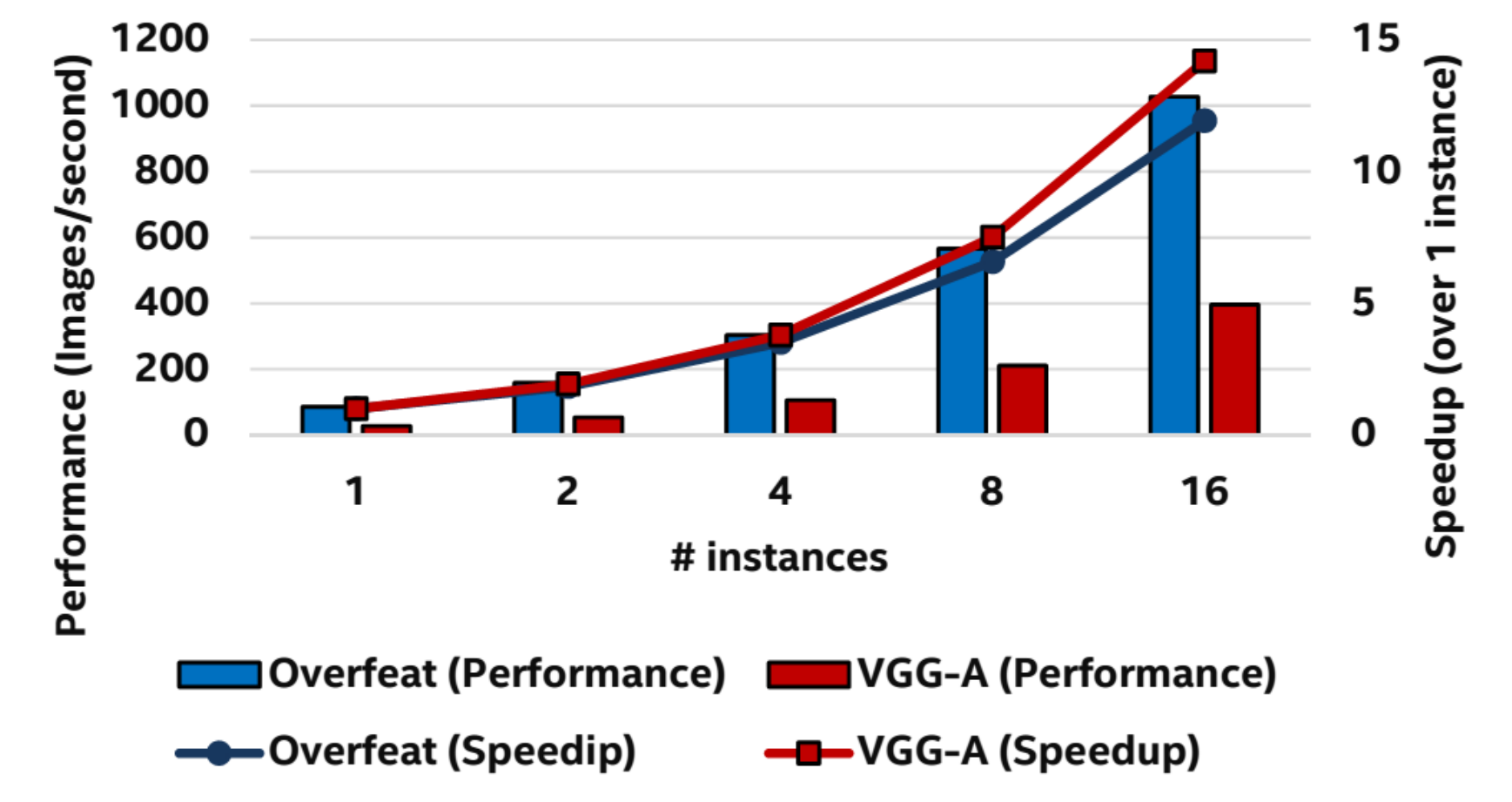}
			\caption{Performance scaling of OverFeat and VGG-A on AWS EC2 with PCL-DNN}
			\label{fig:aws_scaling}
\end{figure}

Figure \ref{fig:aws_scaling} presents the performance (in images/second) and scaling of PCL-DNN on AWS EC2 for Overfeat and VGG-A topologies with mini-batch size of 256. We observe 1027 and 397 images per second on 16 nodes for Overfeat and VGG-A respectively. This translates to 11.9x and 14.2x speedup on 16 nodes respectively. We observe better speedups for VGG-A given its higher  flops per network byte requirements. 

\vspace{-4mm}

\subsection{Automatic Speech Recognition}
\begin{figure}[ht]
     \centering
			\includegraphics[width=.8\columnwidth]{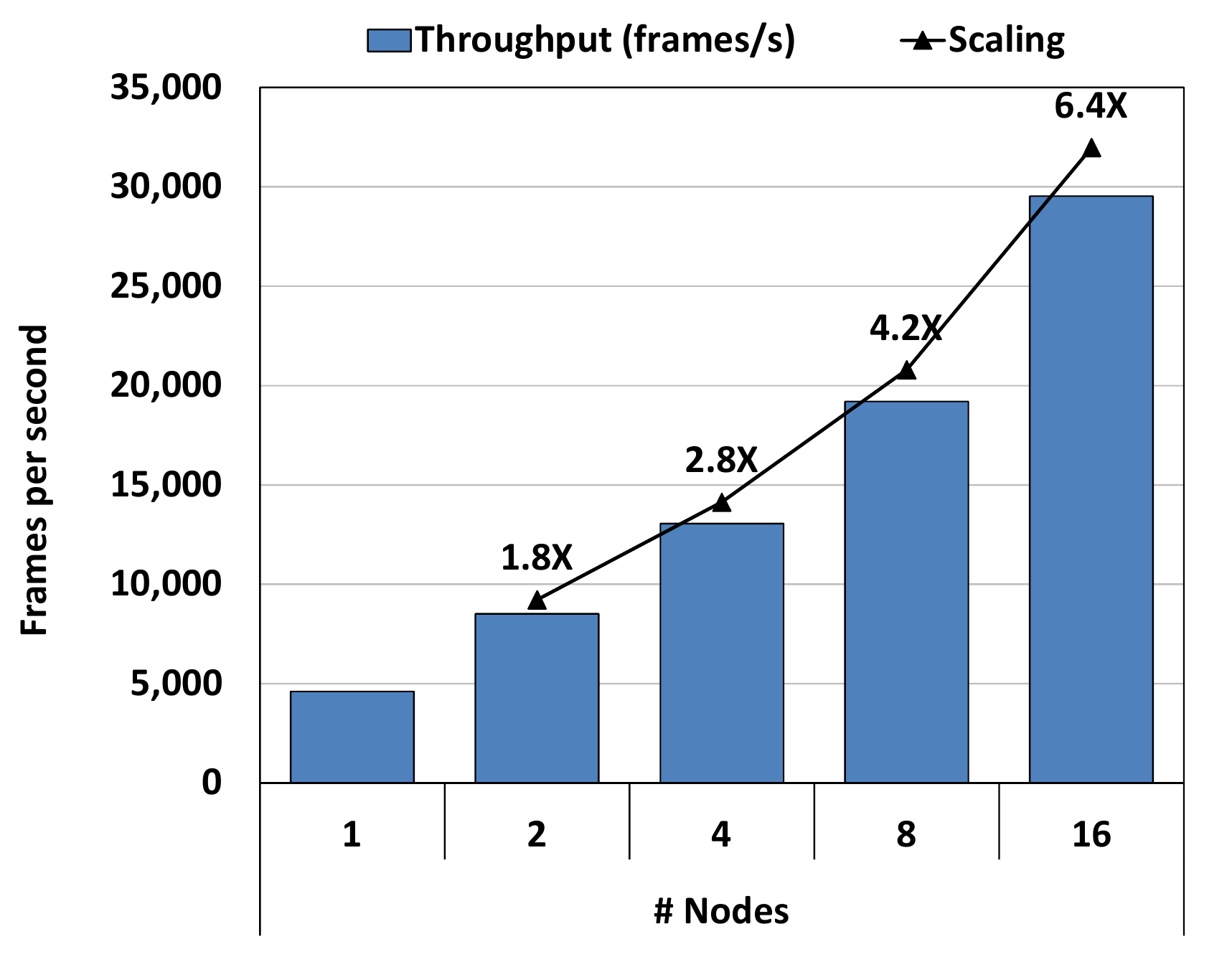}
			\caption{Performance scaling of CD-DNN with PCL\_DNN on the Intel Endeavor cluster upto 16 nodes (each node - Xeon E5-2697v3 HSW CPU with 14x2 cores)}
			\label{fig:cddnn_Scaling}
\end{figure}

We examine DNNs in the Automatic Speech recognition (ASR) context, using Context-Dependent Deep Neural Networks (CD-DNN) HMMs \cite{seide2011}. CD-DNN-HMMs combine classic artificial-neural-network HMMs with traditional tied-state triphones and gives a 33\% relative WER reduction over a discriminatively trained GMM-HMM on Hub5’00 switchboard dataset. This network consists of 7 fully connected hidden layers each with 2048 neurons. For this network a detailed performance study \cite{seide2014}, compares performance for both on single node and scaling.
\\
For the CD-DNN network, on a Xeon E5-2697v3 HSW CPU (with 14x2 cores with 1.7 TFLOPS/s SP peak) PCL-DNN delivers 4600 frames/s. This is $4X$ better than best reported CPU performance and the 2-node performance with PCL-DNN betters the performance reported for an 80-node Xeon cluster \cite{seide2014}. Performance scaling of the CD-DNN network is shown in fig.\ref{fig:cddnn_Scaling}. This surpasses the 3-card K20x performance \cite{seide2014} on an 4 HSW nodes (delivering 13K frames/sec), and continues to scale with a performance of 29.5K frames/sec on 16-nodes.Scaling DNN is far more challenging than the CNNs presented earlier, owing to higher communication to compute ratios and  these results show that PCL-DNN delivers best-in-class performance on CPUs and demonstrating the generality of this optimized framework.

\section{Conclusions}
We demonstrate that deep learning training can be performed at scale using synchronous SGD at high thoughput on CPUs.
We extend the state of the art in terms of scaling and time-to-solution, and present a detailed insights and analysis of multinode training.

\section{Acknowledgements}
The authors would like to acknowledge support from Prabhat Kumar, Wahid Bhimji, and Lisa Gerhardt of the National Energy Research Scientific Computing Center for helping with access and support for the Cori supercomputer.

\newpage
\bibliography{references}

\begin{thebibliography}{13}
\providecommand{\natexlab}[1]{#1}
\providecommand{\url}[1]{\texttt{#1}}
\expandafter\ifx\csname urlstyle\endcsname\relax
  \providecommand{\doi}[1]{doi: #1}\else
  \providecommand{\doi}{doi: \begingroup \urlstyle{rm}\Url}\fi

\bibitem[Abadi et~al.(2015)Abadi, Agarwal, Barham, Brevdo, Chen, Citro,
  Corrado, Davis, Dean, Devin, Ghemawat, Goodfellow, Harp, Irving, Isard, Jia,
  Jozefowicz, Kaiser, Kudlur, Levenberg, Man\'{e}, Monga, Moore, Murray, Olah,
  Schuster, Shlens, Steiner, Sutskever, Talwar, Tucker, Vanhoucke, Vasudevan,
  Vi\'{e}gas, Vinyals, Warden, Wattenberg, Wicke, Yu, and Zheng]{tensorflow}
Abadi, Mart\'{\i}n, Agarwal, Ashish, Barham, Paul, Brevdo, Eugene, Chen,
  Zhifeng, Citro, Craig, Corrado, Greg~S., Davis, Andy, Dean, Jeffrey, Devin,
  Matthieu, Ghemawat, Sanjay, Goodfellow, Ian, Harp, Andrew, Irving, Geoffrey,
  Isard, Michael, Jia, Yangqing, Jozefowicz, Rafal, Kaiser, Lukasz, Kudlur,
  Manjunath, Levenberg, Josh, Man\'{e}, Dan, Monga, Rajat, Moore, Sherry,
  Murray, Derek, Olah, Chris, Schuster, Mike, Shlens, Jonathon, Steiner,
  Benoit, Sutskever, Ilya, Talwar, Kunal, Tucker, Paul, Vanhoucke, Vincent,
  Vasudevan, Vijay, Vi\'{e}gas, Fernanda, Vinyals, Oriol, Warden, Pete,
  Wattenberg, Martin, Wicke, Martin, Yu, Yuan, and Zheng, Xiaoqiang.
\newblock {TensorFlow}: Large-scale machine learning on heterogeneous systems,
  2015.
\newblock URL \url{http://tensorflow.org/}.
\newblock Software available from tensorflow.org.

\bibitem[Chilimbi et~al.(2014)Chilimbi, Suzue, Apacible, and
  Kalyanaraman]{adam}
Chilimbi, Trishul~M., Suzue, Yutaka, Apacible, Johnson, and Kalyanaraman,
  Karthik.
\newblock Project adam: Building an efficient and scalable deep learning
  training system.
\newblock In \emph{11th {USENIX} Symposium on Operating Systems Design and
  Implementation, {OSDI} '14, Broomfield, CO, USA, October 6-8, 2014.}, pp.\
  571--582, 2014.
\newblock URL
  \url{https://www.usenix.org/conference/osdi14/technical-sessions/presentation/chilimbi}.

\bibitem[Deng et~al.(2009)Deng, Dong, Socher, Li, Li, and Fei-Fei]{imagenet}
Deng, J., Dong, W., Socher, R., Li, L.-J., Li, K., and Fei-Fei, L.
\newblock {ImageNet: A Large-Scale Hierarchical Image Database}.
\newblock In \emph{CVPR09}, 2009.

\bibitem[Iandola et~al.(2015)Iandola, Ashraf, Moskewicz, and
  Keutzer]{firecaffe}
Iandola, Forrest~N., Ashraf, Khalid, Moskewicz, Matthew~W., and Keutzer, Kurt.
\newblock Firecaffe: near-linear acceleration of deep neural network training
  on compute clusters.
\newblock \emph{CoRR}, abs/1511.00175, 2015.
\newblock URL \url{http://arxiv.org/abs/1511.00175}.

\bibitem[Schmidhuber(2015)]{Schmidhuber15}
Schmidhuber, J.
\newblock Deep learning in neural networks: An overview.
\newblock \emph{Neural Networks}, 61:\penalty0 85--117, 2015.
\newblock \doi{10.1016/j.neunet.2014.09.003}.
\newblock Published online 2014; based on TR arXiv:1404.7828 [cs.NE].

\bibitem[Seide et~al.(2011)Seide, Li, and Yu]{seide2011}
Seide, Frank, Li, Gang, and Yu, Dong.
\newblock Conversational speech transcription using context-dependent deep
  neural networks.
\newblock In \emph{Interspeech}, pp.\  437--440, 2011.

\bibitem[Seide et~al.(2014{\natexlab{a}})Seide, Fu, Droppo, Li, and
  Yu]{onebitsgd}
Seide, Frank, Fu, Hao, Droppo, Jasha, Li, Gang, and Yu, Dong.
\newblock 1-bit stochastic gradient descent and application to data-parallel
  distributed training of speech dnns.
\newblock In \emph{Interspeech 2014}, September 2014{\natexlab{a}}.
\newblock URL
  \url{http://research.microsoft.com/apps/pubs/default.aspx?id=230137}.

\bibitem[Seide et~al.(2014{\natexlab{b}})Seide, Fu, Droppo, Li, and
  Yu]{seide2014}
Seide, Frank, Fu, Hao, Droppo, Jasha, Li, Gang, and Yu, Dong.
\newblock On parallelizability of stochastic gradient descent for speech dnns.
\newblock In \emph{Acoustics, Speech and Signal Processing (ICASSP), 2014 IEEE
  International Conference on}, pp.\  235--239. IEEE, 2014{\natexlab{b}}.

\bibitem[Sermanet et~al.(2013)Sermanet, Eigen, Zhang, Mathieu, Fergus, and
  LeCun]{overfeat}
Sermanet, Pierre, Eigen, David, Zhang, Xiang, Mathieu, Micha{\"{e}}l, Fergus,
  Rob, and LeCun, Yann.
\newblock Overfeat: Integrated recognition, localization and detection using
  convolutional networks.
\newblock \emph{CoRR}, abs/1312.6229, 2013.
\newblock URL \url{http://arxiv.org/abs/1312.6229}.

\bibitem[Simonyan \& Zisserman(2014)Simonyan and Zisserman]{vgg}
Simonyan, Karen and Zisserman, Andrew.
\newblock Very deep convolutional networks for large-scale image recognition.
\newblock \emph{CoRR}, abs/1409.1556, 2014.
\newblock URL \url{http://arxiv.org/abs/1409.1556}.

\bibitem[Vaidyanathan et~al.(2015)Vaidyanathan, Kalamkar, Pamnany, Hammond,
  Balaji, Das, Park, and Joo]{swoffloadsc2015}
Vaidyanathan, K., Kalamkar, D.~D., Pamnany, K., Hammond, J.~R., Balaji, P.,
  Das, D., Park, J., and Joo, Balint.
\newblock Improving concurrency and asynchrony in multithreaded mpi
  applications using software offloading.
\newblock In \emph{SC15: International Conference for High Performance
  Computing, Networking, Storage and Analysis}, 2015.

\bibitem[Wu et~al.(2015)Wu, Yan, Shan, Dang, and Sun]{deepimage}
Wu, Ren, Yan, Shengen, Shan, Yi, Dang, Qingqing, and Sun, Gang.
\newblock Deep image: Scaling up image recognition.
\newblock \emph{CoRR}, abs/1501.02876, 2015.
\newblock URL \url{http://arxiv.org/abs/1501.02876}.

\bibitem[Zhang et~al.(2014)Zhang, Choromanska, and LeCun]{elasticsgd}
Zhang, Sixin, Choromanska, Anna, and LeCun, Yann.
\newblock Deep learning with elastic averaging {SGD}.
\newblock \emph{CoRR}, abs/1412.6651, 2014.
\newblock URL \url{http://arxiv.org/abs/1412.6651}.

\end{thebibliography}
\bibliographystyle{icml2015}

\end{document}